\documentclass[aps,pra,floatfix,twocolumn,10pt,english,superscriptaddress]{revtex4-1}

\usepackage[utf8]{inputenc}
\usepackage[bitstream-charter]{mathdesign}
\usepackage[T1]{fontenc}

\usepackage{babel}
\usepackage{amsmath}
\usepackage{textcomp}
\usepackage{graphicx}
\usepackage{xcolor}
\usepackage{hyperref}
\hypersetup{
  breaklinks	= true,
  colorlinks	= true,		
  urlcolor	= violet,	
  linkcolor	= blue,		
  citecolor	= violet		
}

\newcommand{\ket}[1]{\vert{#1}\rangle} 
\newcommand{\bra}[1]{\langle{#1}\vert} 
\newcommand{\bracket}[2]{\langle{#1}\vert{#2}\rangle} 
\newcommand{\proj}[1]{\ket{#1}\!\bra{#1}}
\newcommand{\mean}[1]{\langle #1 \rangle}

\newcommand{\one}{\openone}
\DeclareMathOperator{\Tr}{Tr}
\DeclareMathOperator{\e}{e}
\renewcommand{\vec}[1]{\mathbf{#1}}

\newcommand{\abs}[1]{\lvert#1\rvert}

\newcommand{\beq}{\begin{equation}}
\newcommand{\eeq}{\end{equation}}
\newcommand{\up}{\uparrow}
\newcommand{\dn}{\downarrow}
\newcommand{\down}{\downarrow}

\begin{document}

\title{Approaches to Measuring Entanglement in Chemical Magnetometers}

\author{M. Tiersch}
\email{markus.tiersch@oeaw.ac.at}
\affiliation{Institute for Quantum Optics and Quantum Information,
Austrian Academy of Sciences,
Technikerstraße~21A, A-6020 Innsbruck, Austria}
\affiliation{Institute for Theoretical Physics,
University of Innsbruck,
Technikerstraße~25, A-6020 Innsbruck, Austria}

\author{G. G. Guerreschi}
\affiliation{Institute for Quantum Optics and Quantum Information,
Austrian Academy of Sciences,
Technikerstraße~21A, A-6020 Innsbruck, Austria}
\affiliation{Department of Chemistry and Chemical Biology,
Harvard University, Cambridge, Massachusetts 02138 United States}

\author{J. Clausen}
\affiliation{Institute for Quantum Optics and Quantum Information,
Austrian Academy of Sciences,
Technikerstraße~21A, A-6020 Innsbruck, Austria}
\affiliation{Institute for Theoretical Physics,
University of Innsbruck,
Technikerstraße~25, A-6020 Innsbruck, Austria}

\author{H. J. Briegel}
\affiliation{Institute for Quantum Optics and Quantum Information,
Austrian Academy of Sciences,
Technikerstraße~21A, A-6020 Innsbruck, Austria}
\affiliation{Institute for Theoretical Physics,
University of Innsbruck,
Technikerstraße~25, A-6020 Innsbruck, Austria}

\keywords{radical pair mechanism, quantum entanglement, entanglement witness, quantum state tomography, magnetometry}

\begin{abstract}
Chemical magnetometers are radical pair systems such as solutions of pyrene and $N,N$-dimethylaniline (Py--DMA) that show magnetic field effects in their spin dynamics and their fluorescence.
We investigate the existence and decay of quantum entanglement in free geminate Py--DMA radical pairs and discuss how entanglement can be assessed in these systems. We provide an entanglement witness and propose possible observables for experimentally estimating entanglement in radical pair systems with isotropic hyperfine couplings.
As an application, we analyze how the field dependence of the entanglement lifetime in Py--DMA could in principle be used for magnetometry and illustrate the propagation of measurement errors in this approach.
\end{abstract}

\maketitle

\section{Introduction} 
\label{sec:introduction}

Photochemical reactions that involve intermediate radical pairs are known to exhibit magnetic field effects~\cite{Salikhov,Steiner,Nagakura}. The influence of external magnetic fields on these reactions provides a way to use these reactions for measuring and estimating magnetic fields.
For example, the ability of birds and other animals to sense magnetic fields~\cite{magnetoreception,Johnsen,Mouritsen} has been suggested to be based on this spin-chemical mechanism~\cite{Ritz2000,RodgersReview}.
The radical pair mechanism is the model that describes how magnetic field effects arise in these systems~\cite{Salikhov,Steiner,Nagakura}.

Many elements of the radical pair mechanism bear a resemblance to elements in quantum computation procedures or quantum communication protocols. For example, after photoexcitation and charge transfer the initial state of the radical pair is a spin singlet, i.e., a maximally entangled Bell-state, which is a resource state for quantum communication tasks like quantum state teleportation~\cite{Bennett1993,HorodeckiReview}.
The spin state of the radical pair changes due to the presence of the external magnetic field and that of the nuclear spins. Finally, the backward electron transfer completes the chemical reaction by projecting the radical pair spins to the spin singlet state.
In quantum information terminology this projection is known as a Bell-measurement, which also occurs in quantum state teleportation, for example.
These similarities raise the question whether or not magnetic field sensing by means of the radical pair mechanism can also be understood as a simple form of quantum information processing.
A strong indication of whether or not it is \emph{quantum} information processing rather than classical information processing is the presence of quantum entanglement~\cite{HorodeckiReview} between the constituents of the system.

Solutions with two molecular species pyrene (Py) and $N,N$-dimethylaniline (DMA) form radical pairs after a photoexcitation-induced electron transfer and are known to exhibit magnetic field effects~\cite{Schulten,Salikhov,Steiner,Nagakura}. We consider the spin-correlated radical pairs that are formed by one Py and one DMA molecule, in which the spin of the two unpaired electrons is initially in a singlet state. After separation in solution, e.g., by diffusion, the time evolution of the radical pair spins is governed by the strength of the external magnetic field and the hyperfine interaction with nuclear spins of the respective molecule, which we assume to be isotropic due to fast molecular tumbling.
In this situation, the Hamiltonian that generates the dynamics of electron and nuclear spins is given by
\begin{equation} \label{Hamiltonian}
	H = g\mu_{B} \sum_{m=1}^2 \vec{S}_m \cdot \left(\sum_{k=1}^{N_m} \lambda_{mk} \vec{I}_{mk} + \vec{B} \right),
\end{equation}
where the outer sum runs over both molecules of the radical pair and the inner sum is over the $N_m$ nuclei of molecule $m$. The electron spin angular momentum operators are $\hbar\vec{S}_{m}$ and the nuclear spin operators are $\hbar\vec{I}_{mk}$. All nuclear spins are isotropically coupled to the respective electron spin with hyperfine coupling strengths~$\lambda_{mk}$. With the Bohr magneton $\mu_{B}$ and the electron g-factor $g\approx2$ the hyperfine coupling strengths are given in units of millitesla.

Entanglement in radical pair systems has been found in numerical studies of a realistic example of freely diffusing Py--DMA radical pairs~\cite{Cai2010} and radical pair model systems~\cite{Hogben2012}. Here, we revisit entanglement in Py--DMA radical pairs and discuss how entanglement could be experimentally detected in these systems.
Finally, the arising step structure in the magnetic field dependence of the entanglement lifetime in free Py--DMA radicals is analyzed for its suitability for magnetic field measurements.


\section{Entanglement Lifetime of Free Py--DMA Radical Pairs} 
\label{sec:entanglement_lifetime_of_free_py_dma_radical_pairs}

After the creation of the radical pair by photoinduced electron transfer, e.g., Py$^{\bullet-}$ + DMA$^{\bullet+}$, due to the speed of such process, it is a standard assumption that the electron spin state is well described by the singlet state $\rho(0)=\proj{S}$. All the nuclear spins are in the thermal state that, at room temperature, is described by the normalized identity matrix~\cite{Schulten,Salikhov,Steiner,Nagakura}.
After having diffused apart, the exchange and dipolar interaction between the radicals can be neglected and the time evolution of electron and nuclear spins is described by the Hamiltonian~\eqref{Hamiltonian}. Tracing over the nuclear degrees of freedom, the state of the electron spins is then given by
\begin{equation}
	\rho(t) = \Tr_\text{nucl} \left[ U(t) \left(\proj{S}\otimes\frac{\one}{d}\right) U^\dag(t) \right],
\end{equation}
where $U(t)=\exp(-iHt/\hbar)$ and $d$ is the dimension of the nuclear Hilbert space.
It is the electron spin state, which we consider here.
Interactions of the radical pair spins due to re-encounters and the reaction kinetics are not considered in the present treatment, which thus focuses on the spin correlations of geminate free radicals.

The initial singlet state of the two radical pair spins is an entangled state.
A state vector $\ket{\psi}$ of a composite system is called \emph{entangled} if it cannot be written as a product of state vectors of the individual systems, that is, for a composite system formed by subsystems $A$ and $B$ it is not of the form $\ket{\psi}=\ket{\psi_A}\otimes\ket{\psi_B}$. Otherwise $\ket{\psi}$ is called \emph{separable}, that is, not entangled.
For mixed states $\rho$ entanglement is defined by means of decompositions of $\rho$ into convex sums of pure states, e.g., $\rho=\sum_i p_i \proj{\psi_i}$ with probabilities $p_i$ that sum to one. The state $\rho$ is only entangled if it is necessary to use at least one entangled pure state in all of the generally infinite many ways of decomposing $\rho$ into pure states.

To decide whether a given state $\rho$ is entangled is a hard mathematical problem~\cite{HorodeckiReview}, but it has been solved for the case of two spin-\textonehalf{} systems.
Furthermore, the entanglement of such a system can be quantified by an entanglement measure. Such is the concurrence~\cite{Wootters1998} given by $C(\rho) = \max \{0, \sqrt{\lambda_1} - \sqrt{\lambda_2} - \sqrt{\lambda_3} - \sqrt{\lambda_4} \}$, where the $\lambda_i$ are the eigenvalues in decreasing order of the matrix $\rho (\sigma_2\otimes\sigma_2) \rho^* (\sigma_2\otimes\sigma_2)$ with $\sigma_2$ being the second Pauli matrix and $\rho^*$ denoting complex conjugation of the matrix entries in the standard product basis.

Although the initial state of the radical pair spins is the singlet and thus at short times the entanglement in Py--DMA is mainly due to the large singlet contribution to the spin state, the mere presence of coherences in $\rho$ is generally not sufficient for entanglement. For example, the following family of states of two spin-\textonehalf{} particles,
\begin{equation}
	\rho_W(p) = p \proj{S} + \frac{1-p}{4} \one, \qquad 1\ge p \ge -\frac{1}{3},
\end{equation}
contains coherences $\ket{\up\down}\!\bra{\down\up}$ for all $p\ne0$ but it is entangled only for $p>1/3$~\cite{Werner1989}.
A more general consideration leads to further insights into the existence of entanglement as compared to that of coherences. Let us consider all possible quantum states for a given system, which form a continuous convex set of large dimension, e.g., all density operators of two spin-\textonehalf{} particles can be parametrized by 15 real parameters. The subset of states without coherences, i.e., all density matrices that are diagonal in the product basis, is of volume zero within this set, whereas the set of separable (not entangled) states is of finite volume, convex, and centered around the maximally mixed state, which is the density matrix given by the normalized identity matrix. The dynamics of a quantum system given by a time-dependent density operator $\rho(t)$ can be visualized as a continuous curve in the set of states.
Dynamics that take the state asymptotically toward an equilibrium state without coherences will generally exhibit coherences that also only decay asymptotically. This situation can be different when considering entanglement instead. For dynamics that take an initially entangled system asymptotically toward a state that is not entangled and lies within the volume of separable states, there exists a point in time when the curve $\rho(t)$ crosses the boundary between entangled and separable states. That is, at this point in time the state is not entangled any longer, but the dynamics may continue inside the set of separable states leaving the state separable. The disentanglement at finite times, in contrast to an asymptotic decay, is sometimes referred to as ``entanglement sudden death'' in the terminology of the quantum information community~\cite{Eberly}.

\begin{figure}[tb]
	\begin{center}
		\includegraphics[width=\linewidth]{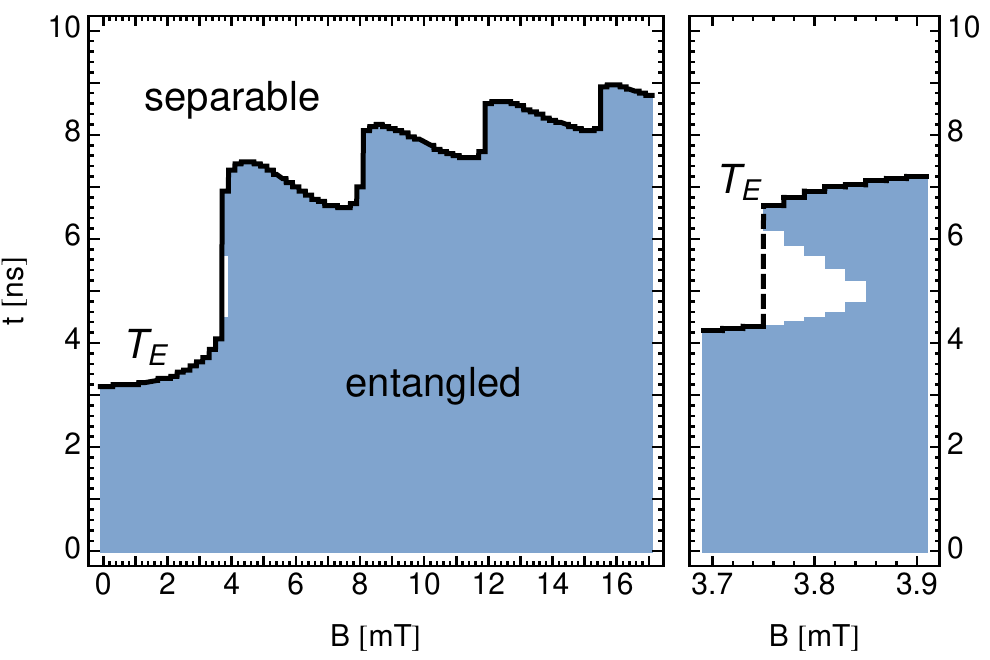}
	\end{center}
	\caption{(Left) Entanglement of the two spin degrees of freedom of geminate free radical pairs as a function of time $t$ and the external magnetic field $B$ for Py-DMA radicals. Hyperfine coupling constants are taken from ref.~\cite{Rodgers2007}. Data points for which the state is entangled are shaded. Data are calculated in steps of $\Delta t=0.04$\,ns and $\Delta B=0.2$\,mT. (Right) Details of entanglement for the first step around 3.8\,mT with $\Delta B=0.02$\,mT where a revival of entanglement occurs. The entanglement lifetime $T_E$ is defined by the last time at which the state is entangled and therefore shows a discontinuity in this region.}
	\label{fig:entanglementLifetime}
\end{figure}

The result of calculating the time-evolution of the free radical pair spins and subsequently testing whether or not the state $\rho(t)$ is entangled is summarized in fig.~\ref{fig:entanglementLifetime} for different strengths of the external magnetic field. This reproduces the findings of the entanglement lifetime for Py--DMA in ref.~\cite{Cai2010} at finer resolution.
Initially, the spins are always entangled because they start out in a singlet state but entanglement decreases in time due to the decoherence introduced by the electrons interacting with the nuclear spin bath~\cite{Tiersch2012}, and vanishes eventually.
The latest time at which entanglement exists defines the entanglement lifetime $T_E = \sup\{t\vert \rho(t) \; \text{entangled}\}$.
As a function of the external magnetic field, $T_E(B)$ shows an increasing trend with several steps.
The overall growth of the entanglement lifetime is caused by the Zeeman shift of the $\ket{\up\up}$ and $\ket{\dn\dn}$ states, which are increasingly separated in energy from the singlet and triplet-zero states. For large $B$ the time evolution of $\rho(t)$ is therefore effectively confined to a smaller dimensional subspace that is spanned by the states $\ket{\up\down}$ and $\ket{\down\up}$. When $\rho(t)$ is fully contained in this subspace the coherences $\ket{\up\down}\!\bra{\down\up}$ that may appear are sufficient for entanglement and thus almost all mixed states in this subspace are entangled.
Another characteristic feature in fig.~\ref{fig:entanglementLifetime} is the steps in the entanglement lifetime. Due to disappearance and revival of entanglement for field strengths around 3.80\,mT (see fig.~\ref{fig:entanglementLifetime} right) the quantity $T_E(B)$ is discontinuous and jumps from $T_E(3.74\,\text{mT})=4.30(2)$\,ns to $T_E(3.76\,\text{mT})=6.62(2)$\,ns.
A revival of entanglement is a hallmark of the non-Markovian nature of the mesoscopic environment of nuclear spins.


\section{Entanglement Witness} 
\label{sec:entanglement_witness}

Entanglement is not an observable but, similar to entropy, for example, is a nonlinear property of the state of two or more quantum systems. In the present case, we consider the entanglement of the two electron spins of the molecules, which form the spin-correlated radical pair, and derive an optimal entanglement witness for radical pairs like Py--DMA.

In general, deciding whether two quantum systems are entangled requires the knowledge of the full density operator of the combined system. Constructing the density operator experimentally via quantum state tomography generically requires the measurement of a tomographically complete set of observables, e.g., all correlation operators $\sigma_i^{(1)}\otimes\sigma_j^{(2)}$ for two spin-\textonehalf{} systems, where $\sigma_i^{(a)}$ is a Pauli matrix or the identity matrix for subsystem $a$.
Observing how entanglement decays in time via state tomography has been undertaken for systems of entangled photons in ref.~\cite{Almeida2007}, for example.

Measuring correlation operators for radical pair systems is challenging because all the different correlation operators cannot be directly measured. Furthermore, rotations of the individual electron spins, $U_1\otimes U_2$, are typically not available in electron spin resonance (ESR) experiments because the two electrons of the radical pair cannot be addressed individually as they can neither be resolved spatially nor in frequency space due to similar $g$-factors.

Although entanglement is not an observable, it is possible to construct observables, so-called entanglement witnesses~\cite{Guehne2009}, from which entanglement can be inferred for some---but not all---quantum states.
Here we define an observable $W$ called an entanglement witness that has expectation values $\mean{W}>0$ for some entangled states and $\mean{W} \le 0$ for all separable states. Note that this definition differs by a sign from the conventional definition~\cite{Guehne2009}.
A measurement outcome $\mean{W}>0$ is only sufficient to demonstrate that a state is entangled, because $\mean{W}\le0$ only allows for the conclusion that the state was either separable or entangled but not detected by the witness.
Therefore, entanglement witnesses are always tailored to specific entangled states.
A witness is optimal for a specific quantum state $\rho$ if $\mean{W}_\rho$ is maximal; i.e., the witness detects all entangled quantum states that lie between $\rho$ and the set of separable states. That is, for an optimal witness there is a family of states for which $\mean{W}_{\rho}>0$ holds if and only if $\rho$ is entangled and, conversely, $\mean{W}_{\rho}\le 0$ implies that a $\rho$ of this family is not entangled.
In contrast to the procedure of a full state tomography, an entanglement witness requires only a single observable to be measured even if it is a collective observable on both subsystems.
Furthermore, an entanglement witness provides a lower bound to the amount of entanglement of the state~\cite{Guhne2007,Eisert2007}, and such a bound can be tightened for a suitable entanglement measure in the case of an \emph{optimal} witness~\cite{Eltschka2012}.

For a general mixed state of two spins $\rho$, a sufficiently large overlap with a maximally entangled state, e.g., the singlet state $\ket{S}=(\ket{\up\dn}-\ket{\dn\up})/\sqrt{2}$, already provides an entanglement witness.
In ref.~\cite{Cai2010} the singlet probability $\bra{S}\rho\ket{S}$ has been proposed as such an entanglement witness for radical pair systems.
It provides a lower bound to the amount of entanglement between the two spins as quantified by the entanglement measure concurrence $C(\rho)$:
\begin{equation}
C(\rho) \ge \max\big\{ 0, 2\bra{S}\rho\ket{S} - 1 \big\},
\end{equation}
that is, a singlet fraction above 1/2 is sufficient to show that the two spins are entangled~\cite{Bennett1996}.
However, a large overlap with the singlet state alone is not necessary for entanglement because the triplet-zero state $\ket{T_0}=(\ket{\up\dn}+\ket{\dn\up})/\sqrt{2}$, which is also often attained by radical pairs, is also a maximally entangled state whereas $\bracket{T_0}{S}=0$.
This observation motivates the construction of an entanglement witness for radical pair systems from a one parameter family of maximally entangled states,
\begin{equation} \label{witness}
W_\phi=2\proj{\phi}-1
\end{equation}
with
\begin{equation}
\ket{\phi}=\left( \ket{\up\down} + \e^{-i\phi}\ket{\down\up} \right) /\sqrt{2},
\end{equation}
which includes as special cases the witness related to the singlet fraction, $W_\pi$, and to the $T_0$-state, $W_0$.
For any separable pure state $\ket{\psi}=(\alpha_1\ket{\dn}+\beta_1\ket{\up})\otimes(\alpha_2\ket{\dn}+\beta_2\ket{\up})$ with normalization $\abs{\alpha_i}^2+\abs{\beta_i}^2=1$ the expectation value of the entanglement witness is $\mean{W_\phi}=\bra{\psi}W_\phi\ket{\psi} \le 0$, which also extends to mixed separable states by linearity.

Systems like Py--DMA exhibit only isotropic hyperfine couplings and hence the total spin of all electrons and nuclei along the direction of the external magnetic field is conserved. Given that the nuclear spins are initially completely depolarized, an initial state with fixed total magnetization of the electrons, e.g., the singlet state, remains under the dynamics generated by the Hamiltonian~\eqref{Hamiltonian} in block-diagonal form in the product basis $\{\ket{\up\up},\ket{\up\down},\ket{\down\up},\ket{\down\down}\}$:
	\begin{equation}
	\label{simplerho}
		\rho(t) = \frac{1}{2} \begin{pmatrix}
		a & 0 & 0 & 0 \\
		0 & b & c & 0 \\
		0 & c^* & b & 0 \\
		0 & 0 & 0 & a
		\end{pmatrix}.
	\end{equation}
The entanglement measure concurrence, evaluated for $\rho(t)$ of this form, yields
\begin{equation} \label{concurrence}
	C(\rho) = \max\{0,\abs{c}-a\}.
\end{equation}
Note that, with the present choice of basis, $\ket{\phi}$ is a column vector taking the form $\ket{\phi}=(0, 1, \e^{-i\phi}, 0)^T/\sqrt{2}$. Representing the matrix element of the spin coherence of $\rho$ as $c=\abs{c}\e^{i\gamma}$, the expectation value of the entanglement witness \eqref{witness} is
\begin{equation}
	\mean{W_\phi}
	= \Tr[\rho W_\phi]
	= \abs{c}\cos(\gamma-\phi) -a.
\end{equation}
For the pertinent entanglement witness for isotropic radical pairs we thus recover that it provides a lower bound to the concurrence for an arbitrary $\phi$, and quantifies concurrence \emph{exactly} for an optimal witness that is tailored to the quantum state with $\phi=\gamma$:
\begin{equation}
	C(\rho) = \max\{0, \mean{W_\gamma} \} \ge \mean{W_\phi}
	\qquad \text{for all $\phi$}.
\end{equation}

To measure the entanglement of $\rho$ exactly by means of this witness, it is necessary to know $\gamma$, which is a parameter of $\rho$, and thus generally time-dependent. A time-resolved measurement of the witness with a fixed $\phi$ gives a lower bound to the entanglement of $\rho(t)$. Following the entanglement dynamics of a time-dependent state~$\rho(t)$ exactly therefore requires some initial knowledge of $\gamma(t)$, which can be obtained as a first guess from a theoretical calculation or by optimization of this angle at each point in time.
Experimentally, it is therefore necessary, in general, to carry out a time-resolved measurement of the time-dependent observable $W_\gamma$.
Note that microscopically the measurement of the witness at different times is done at different radical pair molecules or subensembles, possibly even at different runs of the experiment. We assume, however, that all these molecules are prepared and evolve identically and independent from one another.

Measuring just the entanglement lifetime $T_E$ is simpler because it is only necessary to measure the observable $W_\phi$ with a single constant $\phi$ that is fixed to $\phi=\gamma(T_E)$. The measurement parameter $\phi=\gamma(T_E)$ can either be precomputed from a sufficiently reliable theory or found by experimentally optimizing $\phi$ to give positive measurement results for the latest possible time.


\section{Experimental Considerations} 
\label{sec:experimental_considerations}

Given a system like Py--DMA, i.e., a system with quantum states of the form~\eqref{simplerho}, only the entanglement witness $W_\phi$ with $\phi=\gamma(t)$ needs to be measured to determine the entanglement.
A single observable for measuring entanglement is an improvement over measuring a set of correlation operators and calculating an entanglement measure. However, measuring the witness is still a nontrivial operation on both radical pair molecules.
The entanglement witness can be straightforwardly measured in an experiment only for few choices of $\phi$; e.g., $W_\pi$ is given by the singlet fraction, which is proportional to the singlet fluorescence intensity in Py--DMA systems.
Thus, approaches on how to measure the witness for general $\phi$ or equivalent alternatives are necessary.

A first simplification is to combine several witnesses with fixed angles $\phi$ that promise to be easily obtainable in experiment instead of general time-dependent $\phi$. Would it thus be possible to combine a rather straightforward time-resolved measurement of the singlet fraction with another measurement to obtain the same information as in $\mean{W_\phi}(t)$? After all, the singlet fraction already provides a lower bound on entanglement.
The answer is yes. For example, one can combine measurements of three distinct witnesses with static $\phi=0, \pi/2, \pi$ to replace a measurement with the time-dependent optimal $\phi=\gamma(t)$. These three expectation values for the generic state~\eqref{simplerho} are
\begin{align} 
\mean{W_0} &= + \abs{c}\cos\gamma -a, \label{witness0} \\
\mean{W_\pi} &= -\abs{c}\cos\gamma -a, \\
\mean{W_{\pi/2}} &= - \abs{c}\sin\gamma -a, \label{witnesspi2}
\end{align}
which are essentially given by the $T_0$-fraction, the singlet fraction, and the $T_0$-$S$ coherence, respectively.
These three measurements determine the three real parameters of $\rho$. If one relies on the promise of the special form~\eqref{simplerho} of the density matrix, one can easily evaluate the concurrence~\eqref{concurrence}.
The advantage of the witness is that the corresponding lower bounds on entanglement do not rely on such promise. A better bound, optimal for~\eqref{simplerho} but valid for all states, can be obtained by measuring $W_\gamma$, with $\gamma$ determined by inverting~\eqref{witness0}--\eqref{witnesspi2}.
That is, by measuring these three witnesses one effectively performs state tomography of $\rho$ of the special form~\eqref{simplerho}.

It is conceivable that these three measurements may be obtained with current state-of-the-art techniques such as electron spin resonance (ESR) experiments. There, the electron spin state can be directly addressed with magnetic pulses and different spin components can be observed by applying pulse sequences and measuring the free induction decay. The spin dynamics taking place at time scales of few nanoseconds, however, seem to be at the limit of usual ESR setups.

Another observable that yields information about parameters of the density matrix and may be easier to access in experiment than a generic $W_\phi$ is the total electron spin $\vec{S}_1+\vec{S}_2$. For a single radical pair in state $\rho(t)$ of form \eqref{simplerho} all components of the total spin give
\begin{equation}
\mean{S_1^i+S_2^i} = 0, \qquad i=x,y,z.
\end{equation}
For a single radical pair molecule, all individual outcomes of a measurement of $S_1^i+S_2^i$ are restricted to values -1, 0, and 1. Because $\rho(t)$ is typically not an eigenstate of $S_1^i+S_2^i$ for all times, the fluctuations of the measurement outcomes reveal information about the triplet character of the electron spins, e.g.
\begin{align}
\big[\Delta (S_1^z+S_2^z)\big]^2 &= \mean{(S_1^z+S_2^z)^2} = a, \\
\big[\Delta (S_1^x+S_2^x)\big]^2 &= \mean{(S_1^x+S_2^x)^2} = (1+\abs{c}\cos\gamma)/2.
\end{align}
The coherence between singlet and triplet components, however, cannot be deduced from such polarization measurements.

A measurement that realizes a projection for arbitrary $\phi$ can be realized in principle by a short magnetic pulse parallel to the external magnetic field that inscribes an additional phase difference between the spin-up and down state of one of the radicals, followed by a singlet projection. However, trying to generate such a relevant phase shift on one of the radicals during a typical time span of the spin dynamics ($\sim$1\,ns) by an external pulsed field requires enormous field gradients to generate a sufficient field difference over typical nanometer separation distances of radical pairs in solution. A promising alternative seems to be magnetic nanometer-sized particles, which can supply very localized fields to one of the radicals~\cite{Cai2011}.

\begin{figure}[tb]
	\begin{center}
		\includegraphics[width=\linewidth]{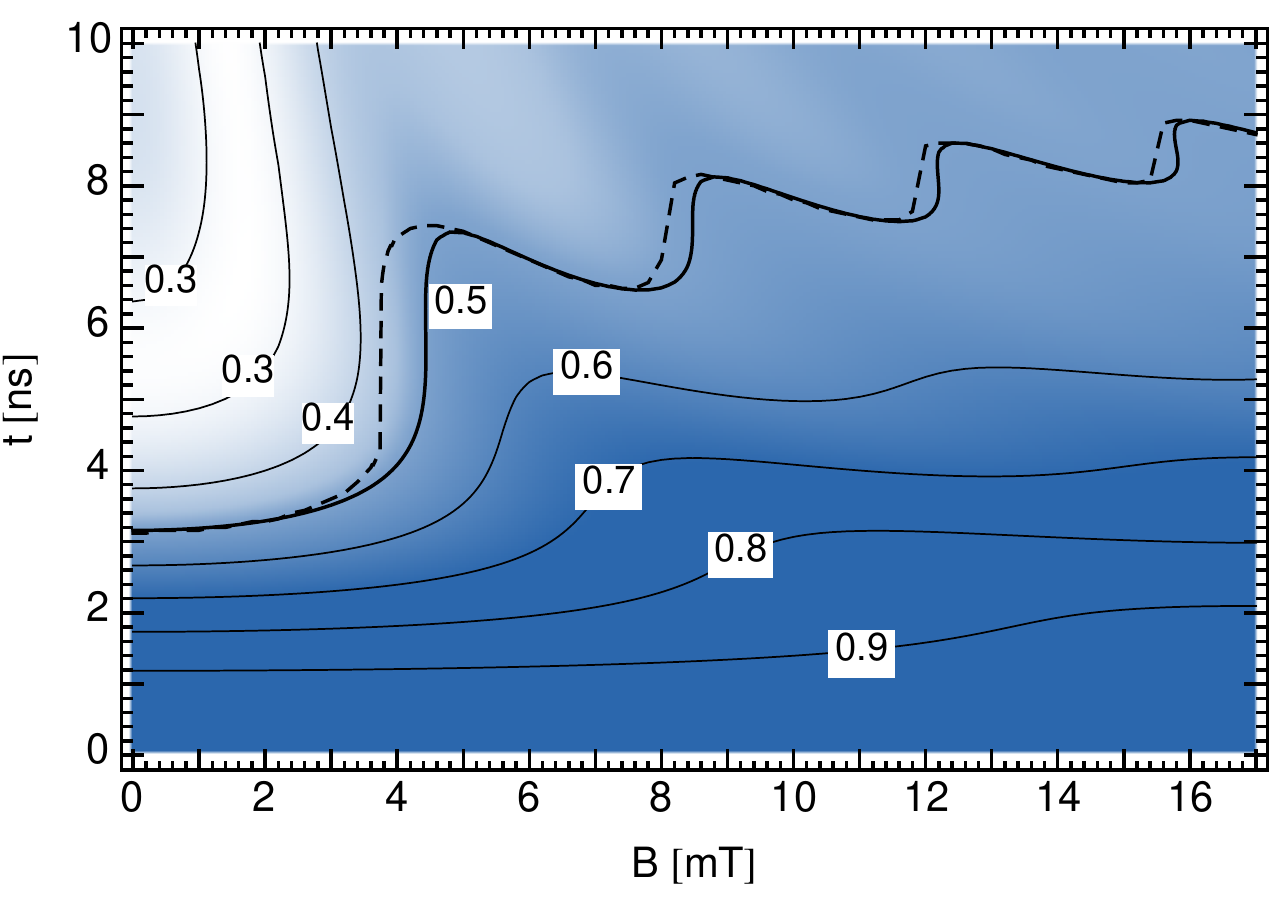}
	\end{center}
	\caption{Time-dependent singlet probability of re-encountering Py--DMA radical pairs for different magnetic fields, depicted by the intensity of the shading in the plot and labeled contour lines, which is an alternative and experimentally more straightforwardly accessible signature. Exciplex fluorescence intensity $I\propto \bra{S}\rho(t)\ket{S}$ above 0.5 gives a lower bound to radical pair entanglement. The contour 0.5  shows similar steep increases in the magnetic field dependence as the entanglement lifetime $T_E$ (dashed).}
	\label{fig:singletfraction}
\end{figure}

Let us finally comment on experimental details of observing just the singlet fraction as a lower bound to entanglement. It can be obtained by measuring the intensity of exciplex fluorescence of re-encountering radical pairs, i.e., by applying a threshold filter to the fluorescence intensity $I\propto \bra{S}\rho(t)\ket{S}$.
The singlet probability for Py--DMA radicals is depicted in fig.~\ref{fig:singletfraction}.

In experimental setups with freely diffusing radical pairs the re-encounter time scale is given by the classical stochastic diffusion process in solution, which is usually modeled as an exponential distribution with a time scale on the order of $\sim 2$\,ns~\cite{Rodgers2007}. That is, geminate radical pairs typically do not exhibit re-encounters at times when the singlet fraction drops below 1/2, but have reacted before. Measurements trying to detect when the singlet fraction drops below 1/2 will therefore suffer from low intensity signals.
The measurement signals can be improved by increasing the probability for the radical pair to re-encounter at later times, e.g., by mounting the radical pair molecules on optically switchable molecules that provide a re-encounter of the radical pairs at a time determined by the experimenter as proposed in ref.~\cite{Guerreschi2013}.
Alternatively, the recombination at longer times can be enhanced by enclosing the radicals in micelles~\cite{Eveson2000} or connecting them with flexible polymer chains~\cite{Weller1984,Enjo1997}. These approaches are experimentally simpler, but lack the additional control offered by molecular optical switches.

For a given re-encounter dynamics, contained in the re-encounter probability distribution $p_\text{re}(t)$, the singlet fluorescence yield until time $t$ is given by
\begin{equation}
	\Phi_S(t) = \int_0^t d\tau\, p_\text{re}(\tau) \bra{S}\rho(\tau)\ket{S}.
\end{equation}
Here we assume that upon a re-encounter the radical pair reacts, thus no longer existing as a radical pair, and that fluorescence occurs immediately upon a re-encounter in a singlet state.
From a time-resolved measurement of the fluorescence intensity, $I(t)\propto d\Phi_{S}(t)/dt$, the singlet fraction can be inferred according to
\begin{equation}
	\bra{S}\rho(t)\ket{S} = \frac{1}{p_\text{re}(t)} \frac{d\Phi_{S}(t)}{dt} \propto \frac{I(t)}{p_\text{re}(t)}
\end{equation}
for known $p_\text{re}(t)$. The constants of proportionality that have been neglected here include the radical pair concentration in solution, excitation (radical pair creation) efficiency, and detection angle, for example. Most of the systematic influences on the intensity can be experimentally determined by a fluorescence measurement at $t=0$ for which $\rho(0)=\proj{S}$ is known.
For estimating the singlet fraction from the fluorescence intensity a precise knowledge of $p_\text{re}$ rather than a phenomenological model is needed, including possible effects of multiple re-encounters~\cite{Clausen2013}, or the circumvention thereof by designing and imposing $p_\text{re}(t)$ experimentally~\cite{Guerreschi2013}.


\section{Magnetic Field Estimation} 
\label{sub:magnetic_field_estimation}

The field dependence of the entanglement lifetime (fig.~\ref{fig:entanglementLifetime}) as discovered in ref.~\cite{Cai2010} exhibits steep increases for some magnetic fields and, due to its definition, even discontinuities. Thus, measurements of the entanglement lifetime can in principle be used to infer or calibrate magnetic fields.

The magnetic field effects on the entanglement lifetime are qualitatively different from usually considered observables such as the singlet yield or radical pair concentration, because the entanglement lifetime is a property of the spin state of the intermediate reactants, in contrast to a time-averaged reaction yield, for example.

At first sight the pronounced field dependence of the entanglement lifetime suggests an extreme sensitivity for magnetic field measurements. For statements about the sensitivity, however, it is necessary to consider the whole process that is required to estimate magnetic fields via the entanglement lifetime.
The time dependence of the entanglement between the two electron spins is a simple consequence of the Hamiltonian~\eqref{Hamiltonian}. The same is true for the entanglement lifetime shown in fig.~\ref{fig:entanglementLifetime}, and there is nothing wrong or unphysical with the sharp field-dependence in this curve, as it was emphatically claimed in a recent paper~\cite{Kominis}.
    
It is an entirely different and independent question how this entanglement, its time evolution, and the time of its disappearance (or ``sudden death''~\cite{Eberly}) due to the hyperfine interaction~\eqref{Hamiltonian} is measured in experiment.  Such an undertaking will comprise at least two tasks.
First, to experimentally access the regime of the plot with large values of the predicted entanglement lifetime (e.g., for values of $B$ larger than 4\,mT), one needs to control the system in such a way that the quantum state of the two electron spins has enough time to evolve before the two radicals reencounter and possibly recombine such that the pair simply vanishes. One possibility, as already discussed in section~\ref{sec:experimental_considerations}, would be to keep the radicals separated in space, e.g., by mounting them on molecular switches (and thus controlling the time of recombination)~\cite{Guerreschi2013}.
Second, it requires a concise description of the procedure how the entanglement is measured, e.g., in terms of witnesses. On the basis of such a description, one can then infer how uncertainties in these measurements translate into uncertainties for its lifetime and thus into precision limits for the magnetic field estimation.
Ignoring these important details may lead to erroneous conclusions regarding the achievable sensitivity in such a hypothetical magnetometer~\cite{Kominis}.

In the following we discuss how limits to the sensitivity of a magnetometer using the entanglement lifetime as a signature arise when taking into account the actual observables that need to be measured.

\subsection{Measurement Errors} 
\label{sub:measurement_errors}

When experimentally determining the entanglement and its lifetime by measuring an optimal entanglement witness, errors of the measurement translate into errors of the inferred entanglement lifetime.
These errors also influence the precision with which a magnetic field could be measured by means of the entanglement lifetime.

Let us consider a time-resolved measurement of the optimal entanglement witness $W_\gamma$. For each measurement time~$t$ one obtains with sufficiently many experimental samples the mean measurement result $\mean{W_\gamma}$ and a confidence region around the mean $[\mean{W_\gamma}-\Delta W_\gamma^{(-)},\mean{W_\gamma}+\Delta W_\gamma^{(+)}]$.
The errors $\Delta W_\gamma^{(\pm)}(t) \ge 0$ to either side of the mean are generally asymmetric, but we omit the additional notation of the superscript $(\pm)$ in what follows.

Starting with an entangled state, the entanglement lifetime $T_E$ is defined as the last time when entanglement exists, that is, afterward $\mean{W_\gamma}(t)\le 0$ for all $t>T_E$.
Due to the experimental uncertainties in the measurement of $W_\gamma$ there will be uncertainties in the last time where $\mean{W_\gamma}(t)$ drops below zero.
These uncertainties given by the confidence interval for $T_E$ can be constructed from the obtained confidence interval of $W_\gamma(t)$ by intersection with the zero line.

The boundaries of the confidence interval for $T_E$ are determined by the first and last time, $T_{E,\text{min}}$ and $T_{E,\text{max}}$ respectively, at which the confidence interval of all $W_\gamma(t)$ includes the zero.
That is, an initially entangled state exhibits $\mean{W_\phi}(t)>0$ for $t<T_E$ and if entanglement can be certified the whole confidence interval takes only positive values for $t\approx 0$. As entanglement decays, the witness finally takes negative values. The confidence interval thus touches the axis $\mean{W_\phi}=0$ for the first time when $\mean{W_\phi}(T_{E,\text{min}})-\Delta W_\phi(T_{E,\text{min}})=0$, which defines $T_{E,\text{min}}=T_E-\Delta T_E$, and for the last time when $\mean{W_\phi}(T_{E,\text{max}})+\Delta W_\phi(T_{E,\text{max}})=0$, which defines $T_{E,\text{max}}=T_E+\Delta T_E$. These intersections define the confidence interval for the entanglement lifetime.
Note, that for large errors $\Delta W_\phi$ it may happen that the zero axis is always included in the confidence region, that is, $\mean{W_\phi}(t)+\Delta W_\phi(t)>0$ for all times $t>T_E$, which means that an upper bound to the entanglement lifetime cannot be given experimentally because it cannot be conclusively shown that entanglement actually disappears.


\subsection{Numerical Example} 
\label{sub:numerical_example}

As an illustration of the analysis, we simulate an experiment that measures the optimal entanglement witness including experimental errors for a freely diffusing Py--DMA radical pair in solution.
We simulate a time-resolved measurement of the entanglement witness $W_\phi$ by doing statistics over 1000 measurements of independently prepared and time-evolved states for each time step $t$ in steps of 0.04\,ns.
To the numerically exact calculation of the spin dynamics for Py--DMA generated by \eqref{Hamiltonian} we add experimental errors for each measurement by introducing small amounts of noise to the exact state. For each time step we analyze the ensemble of 1000 noisy states constructed according to
\[
	\rho(t) = (1-\epsilon) \rho_0(t) + \epsilon \Delta\rho,
\]
where $\rho_0$ is the numerically obtained state evolved according to \eqref{Hamiltonian} and for each measurement the states $\Delta\rho$ are independently sampled uniformly from the state space of mixed states of two spins (Hilbert--Schmidt distributed). This sampling guarantees that the ensemble of $\rho(t)$ remains physical as opposed to simply adding noise to the matrix elements of the exact state.
In total, all error contributions average to the maximally mixed state $\one/4$, which amounts to a fraction $\epsilon$ of white noise added to $\rho_0$.

\begin{figure}[tb]
	\begin{center}
		\includegraphics[width=\linewidth]{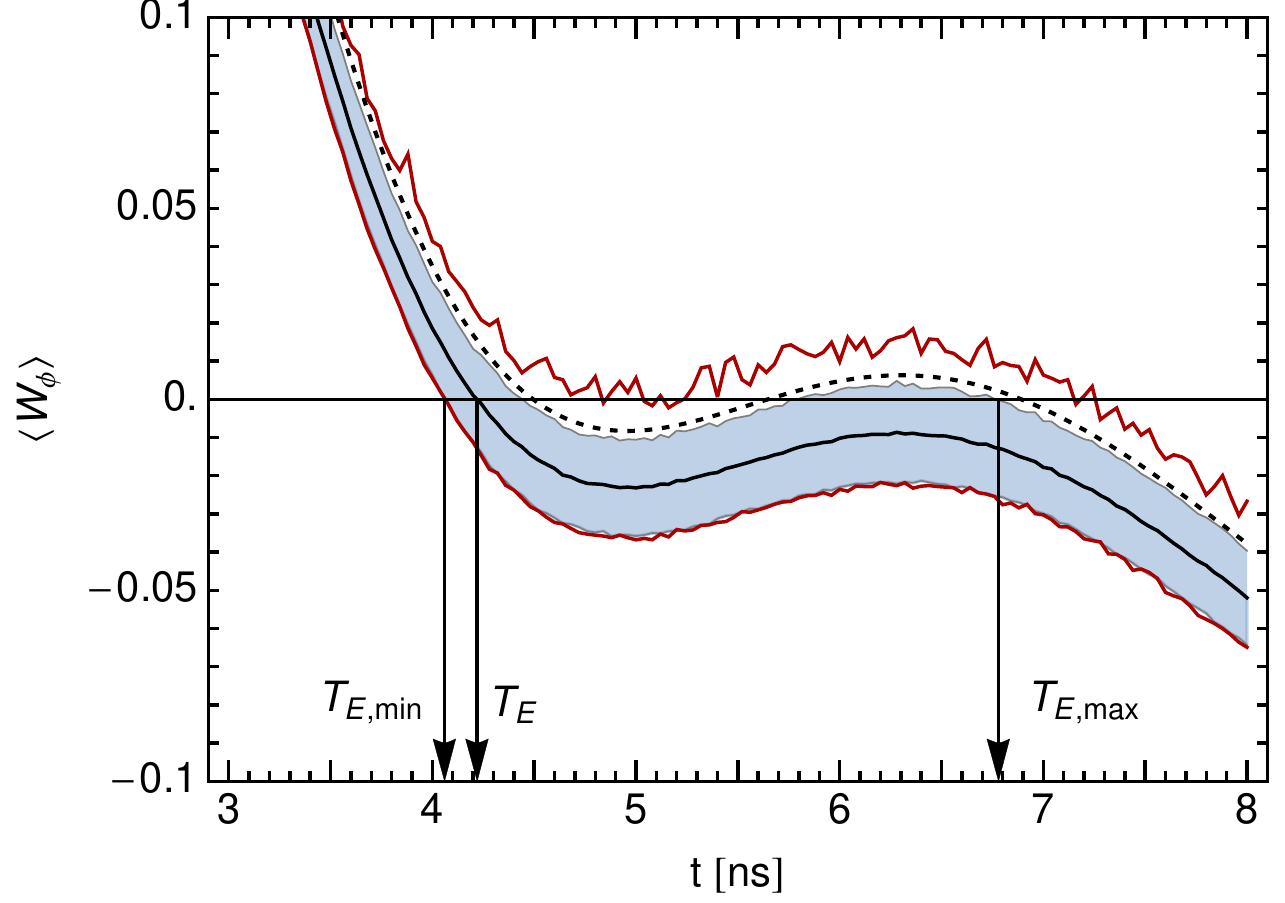}
	\end{center}
	\caption{Confidence interval of a simulated measurement of the optimal entanglement witness at fixed $B=3.8$\,mT in time steps of $\Delta t=0.04$\,ns with $\epsilon=3\,\%$ white noise added to the quantum state. At each time step~$t$, the parameter $\phi$ is chosen such that the mean of the ensemble of 1000 noisy realizations of $\rho(t)$ is maximized. The mean (solid middle line) is surrounded by a $\pm2\sigma$ interval (shaded region), noisy red (outer) lines indicate the minimal and maximal obtained values of the simulated measurements. For comparison the dashed line gives the exact numerical result for $\rho_0(t)$ without noise.}
	\label{fig:figure3}
\end{figure}

Figure~\ref{fig:figure3} shows the time-resolved mean and the confidence interval obtained from the distribution of measurement results of this simulated experiment.
Here, we follow a possible experimental procedure in which the optimal witness parameter is obtained by optimizing $\phi$ at each time step~$t$ independently to give the maximal $\mean{W_\phi}$ for the ensemble. For each time step we generate a sample of 1000 noisy states, for which we optimize $\phi$ to maximize $\mean{W_\phi}$ and calculate the confidence interval.
We find the obtained $\phi$ to coincide with $\gamma$ within numerical and statistical accuracy.
The mean of the obtained distribution coincides with the exact numerical result of $W_\phi$ evaluated for the state $(1-\epsilon)\rho_0(t) + \epsilon \one/4$, i.e., with added white noise, within numerical accuracy.

Within the obtained error bars in fig.~\ref{fig:figure3} there is the possibility of a revival of entanglement because the confidence region intersects with the $\mean{W_\phi}=0$ line twice. For an estimate of the entanglement lifetime and its error bars as observed by the witness, we take the \emph{last} crossing of the mean and the boundaries of the confidence interval of $W_\phi$ with the zero-line.

\begin{figure}[tb]
	\begin{center}
		\includegraphics[width=\linewidth]{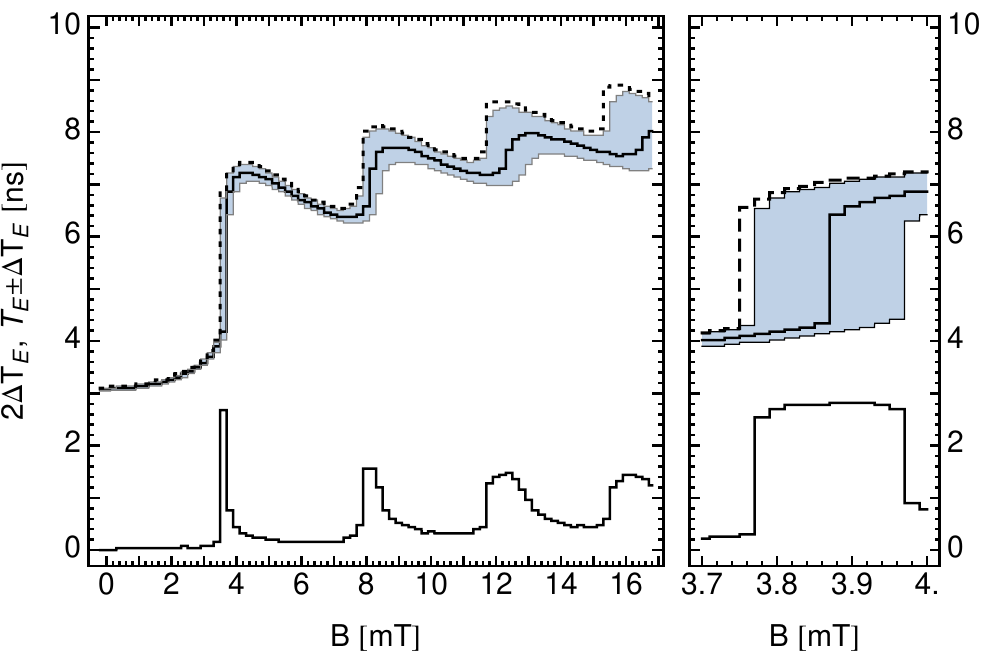}
	\end{center}
	\caption{(Left) Entanglement lifetime as measured in a simulated experiment by means of the optimized entanglement witness with $\epsilon=3\,\%$ white noise (upper curves). The mean (solid) is surrounded by the obtained confidence interval (shaded region). For comparison, the dashed line indicates the entanglement lifetime $T_E$ of the state $\rho_0(t)$ without added noise from fig.~\ref{fig:entanglementLifetime}. The bottom curve quantifies the differences between boundaries of the confidence region, i.e., the (vertical) width of the confidence interval for $T_E$.
	Data are calculated every $\Delta t=0.04$\,ns and $\Delta B=0.2$\,mT.
	(Right) Details of the first jump in entanglement lifetime at 3.87(1)\,mT resolved in steps of $\Delta B = 0.02$\,mT.}
	\label{fig:figure4}
\end{figure}

The resulting magnetic field dependence of the entanglement lifetime and the error bars are shown in fig.~\ref{fig:figure4}. The detailed plot of the first jump in the entanglement lifetime (fig.~\ref{fig:figure4} right) illustrates how the error bars for the entanglement lifetime translate into an error bar of the magnetic field around which the jump occurs.

In order to exploit the steps in the entanglement lifetime experimentally, it is conceivable to use the steps for calibrating the strength of a magnetic field. When increasing the strength of $B$ from a value below the threshold to above the threshold, a sudden spike in the strength of the noise $\Delta T_E$ indicates the region of $B$, where the jump in the entanglement lifetime $T_E$ occurs. The value of $B$ at which the jump occurs cannot be more precisely fixed than the widths of the interval of magnetic fields where the noise is increased.

Note that the error in $B$, as we derived it from errors $\Delta T_E$, depends intricately on the experimentally obtained variation in the prepared quantum states \emph{and} the details of the shape of the curve of the measured witness $W_\phi(t)$.
We expect that a more refined analysis of the confidence interval of the entanglement witness, e.g., including the actual distribution rather than just considering the intersections with the zero line, yields more details in the magnetic field dependence, such as the revival of entanglement and thus smaller inaccuracies in the magnetic field at which the entanglement lifetime jumps.

By scanning over different values of the error $\epsilon$ we observe that larger values of $\epsilon$ generally have the same upper bound of the distribution, but a decreased mean and lower bound for each value of the magnetic field. When adding noise to the states $\rho_0(t)$ the error contributions $\Delta\rho$ are usually biased toward less entangled or separable states. In particular for the initial state $\rho_0(0)=\proj{S}$, which is a maximally entangled state, all errors reduce the entanglement.
For increasingly noisy states the steps at which jumps occur shift toward higher values of $B$ and are increasingly washed out.


In realistic experiments the confidence interval of the entanglement witness is not only given by the errors in the measurement of the witness but also by inaccuracies in the timing, i.e., errors in the time of preparation and measurement, which add errors in the horizontal direction of the curve in fig.~\ref{fig:figure3} and may therefore additionally widen the confidence interval for $T_E$ in fig.~\ref{fig:figure4}.

The spin dynamics in the present scenario is only considered to be generated by the Hamiltonian~\eqref{Hamiltonian}. Typical decoherence sources to the radical pair spin dynamics are dephasing and spin relaxation mechanisms due to fluctuating hyperfine coupling strengths, which, however, happen on time scales of $\sim$1\,\textmu{}s. On short time scales similar to the entanglement lifetime we expect the dominant decoherence mechanism to be caused by stochastic radical pair re-encounters, during which the spin dynamics includes contributions from exchange and dipolar interactions, and influences of the reaction kinematics, which we expect to yield qualitatively similar results (cp.\ figs.~\ref{fig:figure3} and \ref{fig:figure4}).


\section{Conclusion} 
\label{sec:conclusion}

We have revisited the existence and lifetime of entanglement of geminate free Py--DMA radical pairs. Entanglement is a property of the quantum state that, similar to entropy, requires an elaborate method for its experimental detection. In a refined simulation of the spin dynamics, we can identify a revival of entanglement for a magnetic field strength of about 3.8\,mT, which is a clear-cut feature of the non-Markovian dynamics in the mesoscopic spin bath formed by the nuclear spins.

We presented an optimal entanglement witness for entanglement measurements in free radical pair systems, which requires some knowledge about the quantum state or an optimization over its parameter.
The witness applies to all radical pair systems, but it is tailored to radicals that start in typical initial states and evolve under isotropic hyperfine interactions. Possible routes for experimental implementations may be the measurement of three static witnesses that together effectively recover the quantum state of these systems. The information contained in the measurement results of two of these witnesses can also be obtained by measuring the fluctuations of the total radical pair spin parallel and orthogonal to the external magnetic field. Measuring lower bounds on entanglement by means of the singlet product yield have been discussed including the influence of the re-encounter dynamics and reaction kinematics.

Finally, we analyzed the approach to use the entanglement lifetime (fig.~\ref{fig:entanglementLifetime}) as a signature to measure magnetic fields or calibrate certain magnetic field strengths. The magnetic field dependence of the entanglement lifetime and related quantities, which are defined by a threshold, e.g.\ a singlet fraction above 1/2, show a steplike increase with increasing magnetic field strengths. We have illustrated how errors in the primary measurement of entanglement propagate and influence the error in the magnetic field estimation, and we provided a numerical example for Py--DMA radical pairs.
The treatment confirms that, despite the pronounced field dependence of the entanglement lifetime, a physically consistent picture arises once measurement errors are accounted for in detail.
Measuring the entanglement lifetime for magnetometry is admittedly elaborate in comparison to other approaches~\cite{Baker}.
However, when the more general approach is taken and considering also the singlet fraction as a possible experimental signature (fig.~\ref{fig:singletfraction}), which is also easier to access, the magnetic field dependence is qualitatively similar. The singlet fraction above 1/2 does not exhibit a revival occurring in the first step and the steps are located at different values of the magnetic field. It is presently an open question if the field dependence of the entanglement lifetime, the singlet fraction above 1/2, or another similarly defined threshold quantity is best suited for magnetic field estimations.

We note that the scope of the current investigation for radical pairs can also be extended beyond chemistry. For example, there are many formal similarities between radical pairs and quantum dot systems~\cite{Loss} from condensed matter physics (see ref.~\cite{Hanson} for a review) regarding the effective spin Hamiltonian and other environment influences. Quantum dot systems can also exhibit spin--spin entanglement~\cite{Chen,Bayer}, but typically operate in different energy domains and spacial dimensions, which allow for a better control and access to the individual spins by means of additional bias fields.
After the completion of the present work, we learned that similar effects of the entanglement lifetime are also expected in double quantum dot systems~\cite{Mazurek}.
Although spin chemistry experiments generally do not have the same degree of experimental access to the individual spin system as experiments with gated double quantum dot systems, the presented quantum mechanical observables may stimulate further development of ESR methods in this direction.


\begin{acknowledgments}
We acknowledge support by the Austrian Science Fund (FWF) through the SFB FoQuS: F\,4012.
\end{acknowledgments}

\renewcommand{\bibnumfmt}[1]{(#1)}



\begin{thebibliography}{99}

\bibitem{Salikhov}
Salikhov, K. M.; Molin, Y. N. ; Sagdeev, R. Z.; Buchachenko, A. L.
\textit{Spin Polarization and Magnetic Effects in Radical Reactions.}
Elsevier: Amsterdam, The Netherlands, \textbf{1984}

\bibitem{Steiner}
Steiner, U. E.; Ulrich, T.
Magnetic Field Effects in Chemical Kinetics and Related Phenomena.
\textit{Chem.\ Rev.}\ \textbf{1989}, \textit{89}, 51--147

\bibitem{Nagakura}
Nagakura, S.; Hayashi, H.; Azumi, T. (Eds.)
\textit{Dynamic Spin Chemistry. Magnetic Controls and Spin Dynamics of Chemical Reactions.}
Kodansha and Wiley: Tokyo and New York, Japan and U.S.A., \textbf{1998}

\bibitem{magnetoreception}
Wiltschko R.; Wiltschko, W.
Magnetoreception.
\textit{BioEssays} \textbf{2006}, \textit{28}, 157--168

\bibitem{Johnsen}
Johnsen S.; Lohmann, K. J.
The Physics and Neurobiology of Magnetoreception.
\textit{Nat.\ Rev.\ Neurosci.}\ \textbf{2005}, \textit{6}, 703--712

\bibitem{Mouritsen}
Mouritsen, H.; Ritz, T.
Magnetoreception and Its Use in Bird Navigation.
\textit{Curr.\ Opin.\ Neurobiol.}\ \textbf{2005}, \textit{15}, 406--414


\bibitem{Ritz2000}
Ritz, T.; Ademand, S.; Schulten, K.
A Model for Photoreceptor-Based Magnetoreception in Birds.
\textit{Biophys.\ J.} \textbf{2000}, \textit{78}, 707--718

\bibitem{RodgersReview}
Rodgers, C. T.; Hore, P. J.,
Chemical Magnetoreception in Birds: The Radical Pair Mechanism.
\textit{Proc.\ Natl.\ Acad.\ Sci.}\ \textbf{2009}, \textit{106}, 353--360


\bibitem{Bennett1993}
Bennett, C. H.; Brassard, G.; Crépeau, C.; Jozsa, R.; Peres, A.; Wootters, W. K.
Teleporting an Unknown Quantum State Via Dual Classical and Einstein-Podolsky-Rosen Channels.
\textit{Phys.\ Rev.\ Lett.}\ \textbf{1993}, \textit{70}, 1895--1899

\bibitem{HorodeckiReview}
Horodecki, R.; Horodecki, P.; Horodecki, M.; Horodecki, K.
Quantum Entanglement.
\textit{Rev.\ Mod.\ Phys.}\ \textbf{2009}, \textit{81}, 865--942

\bibitem{Schulten}
Werner, H.-J.; Schulten, Z.; Schulten, K.
Theory of the Magnetic Field Modulated Geminate Recombination of Radical Ion Pairs in Polar Solvents: Application to the Pyrene--\textit{N,N}-Dimethylaniline System.
\textit{J. Chem.\ Phys.}\ \textbf{1977}, \textit{67}, 646--663

\bibitem{Cai2010}
Cai, J.; Guerreschi, G. G.; Briegel, H. J.
Quantum Control and Entanglement in a Chemical Compass.
\textit{Phys.\ Rev.\ Lett.}\ \textbf{2010}, \textit{104}, 220502

\bibitem{Hogben2012}
Hogben, H. J.; Biskup, T.; Hore, P. J.
Entanglement and Sources of Magnetic Anisotropy in Radical Pair-Based Avian Magnetoreceptors.
\textit{Phys.\ Rev.\ Lett.}\ \textbf{2012}, \textit{109}, 220501


\bibitem{Wootters1998}
Wootters, W. K.
Entanglement of Formation of an Arbitrary State of Two Qubits.
\textit{Phys.\ Rev.\ Lett.}\ \textbf{1998}, \textit{80}, 2245--2248

\bibitem{Werner1989}
Werner, R. F.
Quantum States With Einstein-Podolski-Rosen Correlations Admitting a Hidden-Variable Model.
\textit{Phys.\ Rev.\ A} \textbf{1998}, \textit{40}, 4277--4281

\bibitem{Eberly}
Yu, T.; Eberly, J. H.
Sudden Death of Entanglement.
\textit{Science} \textbf{2009}, \textit{323}, 598--601


\bibitem{Tiersch2012}
Tiersch, M.; Briegel, H. J.
Decoherence in the Chemical Compass: The Role of Decoherence for Avian Magnetoreception.
\textit{Phil.\ Trans.\ R. Soc.\ A} \textbf{2012}, \textit{370}, 4517--4540

\bibitem{Almeida2007}
Almeida, M. P.; de Melo, F.; Hor-Meyll, M.; Salles, A.; Walborn, S. P.; Ribeiro, P. H. S.; Davidovich, L.
Environment-Induced Sudden Death of Entanglement.
\textit{Science} \textbf{2007}, \textit{316}, 579--582

\bibitem{Guehne2009}
G\"uhne, O.; Tóth, G.
Entanglement Detection.
\textit{Phys.\ Rep.}\ \textbf{2009}, \textit{474} 1--75

\bibitem{Guhne2007}
Gühne, O.; Reimpell, M.; Werner, R. F.
Estimating Entanglement Measures in Experiments.
\textit{Phys. Rev. Lett.}\ \textbf{2007}, \textit{98}, 110502

\bibitem{Eisert2007}
Eisert, J.; Brandão, F. G. S. L.; Audenaert, K. M. R.
Quantitative Entanglement Witnesses.
\textit{New J. Phys.}\ \textbf{2007}, \textit{9}, 46

\bibitem{Eltschka2012}
Eltschka, C.; Siewert, J.
A Quantitative Witness for Greenberger-Horne-Zeilinger Entanglement.
\textit{Sci.\ Rep.}\ \textbf{2012}, \textit{2}, 942


\bibitem{Bennett1996}
Bennett, C. H.; DiVincenzo, D. P.; Smolin, J. A.; Wootters, W. K.
Mixed-State Entanglement and Quantum Error Correction.
\textit{Phys.\ Rev.\ A} \textbf{1996}, \textit{54}, 3824--3851


\bibitem{Cai2011}
Cai, J.
Quantum Probe and Design for a Chemical Compass with Magnetic Nanostructures.
\textit{Phys.\ Rev.\ Lett.}\ \textbf{2011}, \textit{106}, 100501

\bibitem{Rodgers2007}
Rodgers, C. T.; Norman, S. A.; Henbest, K. B.; Timmel, C. R.; Hore, P. J.
Determination of Radical Re-encounter Probability Distributions from Magnetic Field Effects on Reaction Yields.
\textit{J. Am.\ Chem.\ Soc.}\ \textbf{2007}, \textit{129}, 6746--6755


\bibitem{Guerreschi2013}
Guerreschi, G. G.; Tiersch, M.; Steiner, U. E.; Briegel, H. J.
Optical Switching of Radical Pair Conformation Enhances Magnetic Sensitivity.
\textit{Chem.\ Phys.\ Lett.}\ \textbf{2013}, \textit{572}, 106--110

\bibitem{Eveson2000}
Eveson, R. W.; Timmel, C. R.; Brocklehurst, B.; Hore, P. J.; McLauchlan, K. A.
The Effects of Weak Magnetic Fields on Radical Recombination Reactions in Micelles.
\textit{Int.\ J. Radiat.\ Biol.}\ \textbf{2000}, \textit{76} 1509--1522

\bibitem{Weller1984}
Weller, A.; Staerk, H.; Treichel, R.
Magnetic-Field Effects on Geminate Radical-Pair Recombination.
\textit{Faraday Discuss.\ Chem.\ Soc.}\ \textbf{1984}, \textit{78}, 271--278

\bibitem{Enjo1997}
Enjo, K.; Maeda, K.; Murai, H.; Azumi, T.; Tanimoto, Y.
Effect of Polymethylene-Chain Dynamics on the Lifetime of a Charge-Separated Biradical Studied by RYDMR Spectroscopy.
\textit{J. Phys.\ Chem.\ B} \textbf{1997}, \textit{101}, 10661--10665

\bibitem{Clausen2013}
Clausen, J.; Guerreschi, G. G.; Tiersch, M.; Briegel, H. J.
Multiple Re-encounter Approach to Radical Pair Reactions and the Role of Nonlinear Master Equations.
\textit{Preprint} arXiv:1310.6194 (2013)

\bibitem{Kominis}
Kominis, I. K.
Magnetic Sensitivity and Entanglement Dynamics of the Chemical Compass.
\textit{Chem.\ Phys.\ Lett.}\ \textbf{2012}, \textit{542}, 143--146


\bibitem{Baker}
Baker, W. J.; Ambal, K.; Waters, D. P.; Baarda, R.; Morishita, H.; van Schooten, K.; McCamey, D. R.; Lupton, J. M.; Boehme, C.
Robust Absolute Magnetometry with Organic Thin-Film Devices.
\textit{Nat.\ Commun.}\ \textbf{2012}, \textit{3}, 898


\bibitem{Loss}
Loss, D.; DiVincenzo, D. P.
Quantum Computation with Quantum Dots.
\textit{Phys. Rev. A} \textbf{1998}, \textit{57}, 120--126

\bibitem{Hanson}
Hanson, R.; Kouwenhoven, L. P.; Petta, J. R.; Tarucha, S.; Vandersypen, L. M. K.
Spins in Few-Electron Quantum Dots.
\textit{Rev.\ Mod.\ Phys.}\ \textbf{2007}, \textit{79}, 1217--1265

\bibitem{Chen}
Chen, G.; Bonadeo, N. H.; Steel, D. G.; Gammon, D.; Katzer, D. S.; Park, D.; Sham, L. J.
Optically Induced Entanglement of Excitons in a Single Quantum Dot.
\textit{Science} \textbf{2000}, \textit{289}, 1906--1909

\bibitem{Bayer}
Bayer, M.; Hawrylak, P.; Hinzer, K.; Fafard, S.; Korkusinski, M.; Wasilewski, Z. R.; Stern, O.; Forchel, A.
Coupling and Entangling of Quantum States in Quantum Dot Molecules.
\textit{Science} \textbf{2001}, \textit{291}, 451--453


\bibitem{Mazurek}
Mazurek, P.; Roszak, K.; Chhajlany, R. W.; Horodecki, P.
Sensitivity of the Decay of Entanglement of Quantum Dot Spin Qubits to the Magnetic Field.
\textit{Preprint} arXiv:1304.1749 (2013)


\end{thebibliography}
\end{document}